\documentclass[11pt]{amsart}
\usepackage{geometry}                
\usepackage{graphicx}
\usepackage{amssymb}
\usepackage{epstopdf}
\title{Lempel-Ziv complexity reference}
\author{G. Ruffini - Jan 2017 --- Starlab Technical Note, TN00344 (v1.0)}

\usepackage{color}

\usepackage{listings}             

\begin{document}

\maketitle
\tableofcontents
\section*{Abstract}
The aim of this note is to  provide some reference facts for LZW---mostly from Thomas and Cover \cite{Cover:2006aa}---adapted to the needs of the  Luminous project. LZW is an algorithm to compute a Kolmogorov Complexity estimate derived from a limited programming language that only allows copy and insertion in strings (not Turing complete set). 

Despite its delightful simplicity, it is rather powerful and fast. We then focus on definitions of LZW derived complexity metrics consistent with the notion of descriptive length, and discuss different normalizations, which result in a set of metrics we call $\rho_0$, $\rho_1$ and $\rho_2$, in addition to the Description Length $l_{LZW}$ and the Entropy Rate.


\clearpage

\section{LZW compression: the main concept}

The main idea in LZW is to look for repeating patterns in the data, and instead of rewriting repeating sequences, refer to the last one seen \cite{Lempel:1976aa}.  As Kaspar clearly states, LZW is the Kolmogorov Complexity computed with a limited set of programs that only allow copy and insertion in strings \cite{Kaspar:1987aa, Ruffini:2016ad}. \\

``We do not profess to offer a new absolute measure for complexity which, as mentioned already, we believe to be nonexistent. Rather, we propose to evaluate the complexity of a finite sequence from the point of view of a simple self- delimiting learning machine which, as it scans a given n- digit sequence $S=s_{1}\cdot s_{1} \cdot ... s_{n}$ , from left to right, adds a new word to its memory every time it discovers a substring of consecutive digits not previously encountered.  The size of the vocabulary, and the rate at which new words are encountered along $S$, serve as the basic ingredients in the proposed evaluation of the complexity of $S$.''\\

We consider a string of characters in and alphabet with $A$ symbols (typically binary) of length $n$. 
From wikipedia:  A high level view of the encoding algorithm is shown here:

 {\footnotesize
\begin{lstlisting}%[frame=single]  % Start your code-block
%P = [set of parameters]
%B= [set of backgrounds]
1. Initialize the dictionary to contain all strings of length one.
2. Find the longest string W in the dictionary that matches the current input.
3. Emit the dictionary index for W to output and remove W from the input.
4. Add W followed by the next symbol in the input to the dictionary.
5. Go to Step 2.
\end{lstlisting}
}

After applying LZW, we will end up with a set of words (or phrases, as they are sometimes called)  $c(n)$ that go into a dictionary. The length of the compressed string will be $l_{LZW} \leq n$ (the analog of Kolmogorov or algorithmic complexity).  \\

 The {\bf description length of the sequence encoded by LZW}  would have length less or equal to the number of phrases times the number of bits needed to identify a seen phrase plus the bits to specify a new symbol (to form a new phrase), hence\footnote{
 Actually, we can do a bit better than this. In practice, not all dictionary entries are used. We can use the max dictionary key ID and state that ``n bits are needed to describe any key entry, and there are m of them (and here they are)", leading to 
$
l_{LZW} \le \log_{2}(\log_{2}  \max(output))+ \mbox{length}(output) *  \log_{2} \left[  \max(output) \right]
$, 
since we need $ \log_{2}(\log_{2}  \max(output))$  bits to describe $n$. This is how it is implemented in the appended code.
}
\begin{equation}
l_{LZW} \le c(n) \log_{2} \left[ c(n)+ \log_{2} A \right] \approx c(n) \log_{2} \left[ c(n)\right]
\end{equation}


\section{The process of digitization}
When we digitize (e.g., binarize) a signal prior LZW, we are creating a new string from the data, and we make an explicit choice on what aspects of the data we wish to compress.   In this process we destroy information---we are going to do lossy compression. Thus, the choice of digitization results in us having access to a subset of the features of the original string. \\

A reasonable strategy is to preserve as much information as possible in the resulting transformed string. In this sense, using methods that maximize the entropy of the resulting series are recommended, such as using the median for thresholding (this is guaranteed to result  in $H_{0}=1$)\footnote{Can we generalize this idea? Can we, e.g., binarize the data so that it has maximal $H_{0}$ and $H_{1}$?}.  
\\

On the other hand, other methods that destroy more information may tap  and highlight  other, also relevant features of the data.  At this stage, then, how to binarize or preprocess (e.g., filter) the original string  is an empirical question. The same applies to the choice of compression method, of course, as LZW is just one framework for compression.


\section{LZW and entropy rate for stochastic processes}

The main fact from Thomas and Cover \cite{Cover:2006aa} refers to stochastic random processes $\{X_{i}\}$. A key concept is the {\bf entropy rate} of the stochastic process, given by  
 \begin{equation}
 \mathcal H(X)= \lim_{n\rightarrow \infty} {1\over n} H(X_1, ..., X_n), 
\end{equation}
when this limit exists, with $H$ denoting the usual multivariate entropy of $X$, $
H(X)=-E_{X}[\log(P(X)]
$.  It is an important theorem that for stationary processes, 
\begin{equation}
 \mathcal H(X)
= \lim_{n\rightarrow \infty} H(X_n|X_{n-1}, X_{n-2} ..., X_1).
\end{equation} 
\newline 

Let also  
$$H_{0}(p) = -p\log p -(1-p)\log (1-p)$$ 
denote the {\bf univariate entropy}, with $p$ the probability of a Bernoulli (binary) process (Markov chain\footnote{We denote a {\bf Markov chain of order $m$} to be one where the future state depends on the past $m$ states (time-translation invariantly), $P(X_{n}|X_{n-1}, ..., X_{1})= P(X_{n}|X_{n-1}, ..., X_{n-m})$ for $n>m$. } of order zero).  \\

We note that entropy rate of a stochastic processes is non-increasing as a function of order, that is, $0\leq \mathcal H \leq .. \leq H_{q} \leq ... \leq H_{0} \leq 1$.  \\

The fundamental relation is that description length is closely related to entropy rate, 
\begin{equation}
l_{LZW}= c(n) \log_{2} \left[ c(n)+ \log_{2} A \right] \approx c(n) \log_{2} \left[ c(n)\right] \longrightarrow {n}{\mathcal H} 
\end{equation}
\newline

Another important  result  in what follows is that with probability 1 (Thomas and Cover Theorem 13.5.3)
$$
\lim_{n\rightarrow \infty} \sup   l_{LZW} \leq n \mathcal H 
 $$ 
 which can rewrite as
 $$
 \lim_{n\rightarrow \infty} \sup   c(n) \log_{2} c(n) \leq  n \mathcal H  \leq n  H_{0}
 $$
 and use to rewrite (in the limit above)
\begin{equation}
 c(n)  \leq  \frac{n \mathcal H}{\log_{2} c(n) }  \leq \frac{n \mathcal H}{\log_{2} \frac{n \mathcal H}{\log_{2} c(n) }}  \sim \frac{n \mathcal H}{\log_{2} n } 
 \leq \frac{n  H_{0}}{\log_{2} n }  
 \end{equation}
 which we use below for normalization purposes.
\section{Metrics}

Two metrics are used in the field, one is $c(n)$ and the other $l_{LZW}$. Of the two the latter is more closely related to Kolmogorov complexity or description length. Both contain similar information (in fact one is a monotonic function of the other).

\section{Fundamental Normalization of LZW}

The purest way to normalize this metric is to normalize  by the original string length $n$
$$
\rho_{0} = l_{LZW} / n = \frac{c(n) \log_{2} [ c(n) +A] } {n} \rightarrow  \mathcal H
$$
with units of bits per character. This is the  {\bf  LZW compression ratio}.

\section{Other normalizations or measures}

A typical normalization adopted by  the literature is to ``divide by entropy''. By this we mean $\rho_{1}= l_{LZW} / H_{0}$. In the literature this is usually defined through $c(n)$, 
$$
\rho_{1} = \frac{c(n)}{\frac{n  H_{0}}{\log_{2} n } } \sim \frac{\mathcal H}{H_{0}} \sim \frac{l_{LZW} }{nH_{0}} \rightarrow \frac{\mathcal H}{H_{0}}
$$
(with units of bits per character).
Essentially the same can be computed from the randomly reshuffled data series, which with high probability forces $l_{LZW} \sim n H_{0}$ by destroying 2nd order interactions.
Hence,
$$
\rho_{1} \approx  \frac{l_{LZW}} { H_{0}} \approx \frac{l_{LZW} }{l^{shuf}_{LZW}} 
$$
This ratio tells us how much information density is hidden in 2nd and higher order entropy rate as compared to first order one. \\

We can think of this a being the comparison of ``first order apparent complexity'' (entropy) and an estimate of the entropy rate (which provides and upper bound to algorithmic complexity). This is an important comparison, as the proposal in \cite{Ruffini:2016ac} is that conscious level may be associated to systems that exhibit high apparent entropy with low algorithmic complexity.

Alternatively, we could define
$$
\rho_{2}= H_{0} -  \rho_{0} >0
$$
which can be interpreted as the extra apparent extra entropy (bits/char) incurred by using first order methods instead estimating the true entropy rate. \\

At any rate, from a machine learning point of view it is probably best to compute $\rho_{0}$, $H_{0}$, ..., $H_{q}$ as separate measures.
\\

Also, it is known that LZW or entropy estimates are sensitive to string length. When comparing metrics across datasets make sure you keep string length constant and as long as possible.

\section{LZW and Kolmogorov complexity}
As mentioned above, LZW description length is only an estimate of algorithmic complexity. We can easily provide examples of sequences that have low algorithmic complexity, but which are rather hard to compress using LZW. For example, consider the firs $n$ digits of $\pi$. Or consider ``the digits of the smallest prime with $10^9$ digits''. The algorithmic complexity of these numbers is very low, but LZW won't compress them much if at all. Does this mean that LZW is useless? No, but we should keep in mind that it provides only {\bf an upper bound on algorithmic complexity.} In order to get better bounds, we may consider a family of compressors and use the lowest complexity that any of them can find as a better upper bound. 

\footnotesize
\bibliographystyle{plain} 
\bibliography{kolmogorov}  

\begin{thebibliography}{1}

\bibitem{Cover:2006aa}
Thomas~M. Cover and Joy~A. Thomas.
\newblock {\em Elements of information theory}.
\newblock John Wiley \& sons, 2 edition, 2006.

\bibitem{Kaspar:1987aa}
F.~Kaspar and H.~G. Schuster.
\newblock Easily calculable measure for the complexity of spatiotemporal
  patterns.
\newblock {\em Phys. Rev. A}, 36(2):842--848, 1987.

\bibitem{Lempel:1976aa}
A~Lempel and J~Ziv.
\newblock On the complexity of finite sequences.
\newblock {\em IEEE Transactions on Information Theory}, IT-22(1):75--81,
  January 1976.

\bibitem{Ruffini:2016ac}
G.~Ruffini.
\newblock An algorithmic information theory of consciousness.
\newblock {\em Submitted to the Neuroscience of Consciousness}, August 2016.

\bibitem{Ruffini:2016ad}
Giulio Ruffini, David~Iba\ nez, Marta Castellano, Stephen Dunne, and Aureli
  Soria-Frisch.
\newblock Eeg-driven rnn classification for prognosis of neurodegeneration in
  at-risk patients.
\newblock {\em ICANN 2016}, 2016.

\end{thebibliography}
\clearpage

\definecolor{javared}{rgb}{0.6,0,0} 
\definecolor{javagreen}{rgb}{0.25,0.5,0.35} 
\definecolor{javapurple}{rgb}{0.5,0,0.35} 
\definecolor{javadocblue}{rgb}{0.25,0.35,0.75} 
 
\lstset{language=Python,
basicstyle=\ttfamily\tiny,
keywordstyle=\color{blue}\bfseries,
stringstyle=\color{red},
commentstyle=\color{javagreen},
morecomment=[s][\color{javadocblue}]{/**}{*/},
numbers=left,
numberstyle=\tiny\color{black},
stepnumber=1,
numbersep=10pt,
tabsize=4,
showspaces=false,
showstringspaces=false}

{\scriptsize
\section*{Annex: StarLZW.py}
\begin{lstlisting}[language=Python]
# LZW Update May 2017, Feb 12 2017 to match TN00344. 
# G. Ruffini with Eleni Kroupi / Starlab + Neuroelectrics (2017)
# Updated May 2017 with Entropy rate calculators
# Part of this code from sources as indicated, but modified to 
# be used with binary data.

# Permission is hereby granted, free of charge, to any person obtaining a copy of 
# this software and associated documentation files (the "Software"), to deal in 
# the Software without restriction, including without limitation the rights to 
# use, copy, modify, merge, publish, distribute, sublicense, and/or sell copies of
# the Software, and to permit persons to whom the Software is furnished to do so, 
# subject to the following conditions:
# 
# The above copyright notice and this permission notice shall be included in all 
# copies or substantial portions of the Software.
# 
# THE SOFTWARE IS PROVIDED "AS IS", WITHOUT WARRANTY OF ANY KIND, EXPRESS OR 
# IMPLIED, INCLUDING BUT NOT LIMITED TO THE WARRANTIES OF MERCHANTABILITY, FITNESS
# FOR A PARTICULAR PURPOSE AND NONINFRINGEMENT. IN NO EVENT SHALL THE AUTHORS OR 
# COPYRIGHT HOLDERS BE LIABLE FOR ANY CLAIM, DAMAGES OR OTHER LIABILITY, WHETHER 
# IN AN ACTION OF CONTRACT, TORT OR OTHERWISE, ARISING FROM, OUT OF OR IN 
# # CONNECTION WITH THE SOFTWARE OR THE USE OR OTHER DEALINGS IN THE SOFTWARE.      



import numpy as np
##########################################################################

def binarizeby(data, method="median"):
    """ Binarize 1D numpy data array using different methods: mean or median"""
    
    d=data*0
    if method == "median":
        thr=np.median(data)
        d[np.where(data>= thr)]=1
        print " Using median as threshold"
        print " The median is:", str(thr)
                                    
    else: # use mean
        thr=np.mean(data)
        d[np.where(data>= thr)]=1
        print " Using mean as threshold"
        print " The mean is:", str(thr)
    print " Binarized data average", str(np.mean(d.astype(np.int8)))
    return d.astype(np.int8)


# adapted from https://rosettacode.org/wiki/LZW_compression#Python
def compress(theString,mode='binary',verbose=True):
    """Compress a *string* to a list of output symbols. Starts from two sympols, 0 and 1. 
       Returns the compressed string and the length of the dictionary
       If you need to, convert first arrays to a string , 
       e.g., entry="".join([np.str(np.int(x)) for x in theArray]) """
 
    if mode == 'binary':
        dict_size=2
        dictionary={'0':0, '1':1}
    elif mode=='ascii':
        # Build the dictionary for generic ascii.
        dict_size = 256
        dictionary = dict((chr(i), i) for i in xrange(dict_size))
    else:
        print "unrecognized mode, please use binary or ascii"
    w = ""
    result = []
    for c in theString:
        wc = w + c

        if wc in dictionary:
            w = wc
        else:
            result.append(dictionary[w])
            # Add wc to the dictionary.
            dictionary[wc] = dict_size
            dict_size += 1
            w = c
 
    # Output the code for w.
    if w:
        result.append(dictionary[w])
    if verbose:    
        print "length of input string:", len(theString)    
        print "length of dictionary:", len(dictionary)
        print "length of result:", len(result)
    return result,len(dictionary)



def decompress(compressed,mode='binary'):
    """Decompress a list of output ks to a string."""
    from cStringIO import StringIO
 
    if mode == 'binary':
        dict_size=2
        dictionary={0:'0', 1:'1'}
        w = str(compressed.pop(0))


    elif mode=='ascii':
        # Build the dictionary for generic ascii.
        dict_size = 256
        dictionary = dict((i, chr(i)) for i in xrange(dict_size))
        w = chr(compressed.pop(0))

    else:
        print "unrecognized mode, please use binary or ascii"

    # use StringIO, otherwise this becomes O(N^2)
    # due to string concatenation in a loop
    result = StringIO()
    result.write(w)

    for k in compressed:
        if k in dictionary:
            entry = dictionary[k]
        elif k == dict_size:
            entry = w + w[0]
        else:
            raise ValueError('Bad compressed k: %s' % k)
        result.write(entry)
 
        # Add w+entry[0] to the dictionary.
        dictionary[dict_size] = w + entry[0]
        dict_size += 1
 
        w = entry
    return result.getvalue()
      
    
    
############################################################################## 
                      
def Compute_rho0(theArray):
    """Computes rho0 metric (bits/Sample) as described in TN000344 Starlab / Luminous """

    return    ComputeDescriptionLength(theArray) / len(theArray)*1.0   

def Compute_rho1(theArray):
    """Computes rho1 metric (bits/Sample) as described in TN000344 /Starlab Luminous"""

    return    ComputeDescriptionLength(theArray) / ShannonEntropy(theArray)  / len(theArray)

def Compute_rho2(theArray):
    """Computes rho2 metric (bits/Sample) as described in TN00044 """

    return    Compute_rho0(theArray)- ShannonEntropy(theArray)  


def ComputeDescriptionLength(theArray,classic=False):
    """Computes description lenght l_{LZW} as described in TN000344"""

    entry="".join([np.str(np.int(x)) for x in theArray])   
    
    compressedstring,len_dict=compress(entry)  # returns 
    ndigits=len(np.unique(theArray))
    print "distinct digits:", ndigits
    

    
    # you need n bits log(max(comp))) to describe this sequence. And there are this many: len(a)...
    # DL: "you need n bits. There are these (m) of them".

    DL=np.log2(np.log2(max(compressedstring)))+ np.log2(max(compressedstring))*len(compressedstring)
    
    if classic:
    # old way ... more for LZ than LZW:
      DL= len_dict*np.log2(len_dict)
    
    return DL

def ShannonEntropy(labels):
    """ Computes entropy of label distribution. numpy array int in """
    n_labels = len(labels)
    if n_labels <= 1:
        return 0

    counts = np.bincount(labels)
    probs = counts*1.0 / n_labels*1.0
    n_classes = np.count_nonzero(probs)
  
    if n_classes <= 1:
        return 0
    
    # Compute standard entropy.
    ent = 0.
    for i in probs:
        ent -= i * np.log2(i)

    return ent


##########################################################################
# Entropy rate of English ...

import re, codecs, random, math, textwrap
from collections import defaultdict, deque, Counter

def ComputeER(theArray, markov_order=4):
        """ Compute entropy rate of array with some memomory / markov_order"""
        # compute entropy rate ...
        #np.savetxt("rule.txt",  theArray.astype(int),fmt='%i', delimiter="")
        #model, stats = markov_model(chars("rule.txt"), markov_order)
        datastream=streamArray(theArray.astype(int))
        model, stats = markov_model(datastream, markov_order)
        del datastream
        return entropy_rate(model, stats)
    
    
def streamArray(theArray):
    """Generator for array elements (convenient)"""
    for element in theArray:
        yield element
        

# The following 3 are adapted from:  
# Clement Pit--Claudel (http://pit-claudel.fr/clement/blog)
def markov_model(stream, model_order):
    """model describes prob of a digit appearing following n prior ones"""
    """ Stats provide frequencies of groups of n symbols"""
    model, stats = defaultdict(Counter), Counter()
    circular_buffer = deque(maxlen = model_order)
    
    for token in stream:
        prefix = tuple(circular_buffer)
        circular_buffer.append(token)
        if len(prefix) == model_order:
            stats[prefix] += 1.0
            model[prefix][token] += 1.0
    return model, stats

def entropy(stats, normalization_factor):
    return -sum(proba / normalization_factor * math.log(proba 
                    / normalization_factor, 2) for proba in stats.values())

def entropy_rate(model, stats):

    return sum(stats[prefix] * entropy(model[prefix], stats[prefix]) 
               for prefix in stats) / sum(stats.values())




    
\end{lstlisting}
}

\end{document}